\documentstyle[epsfig,aps]{revtex}
\begin{document}

\title{The Bak-Sneppen Model on Scale-Free Networks}

\author{Yamir Moreno$^1$\cite{byline}, Alexei Vazquez$^2$}

\address{$^1$ The Abdus Salam International Centre for
Theoretical Physics,\\ Condensed Matter Group, P.O. Box 586, Trieste,
I-34014, Italy.\\ $^2$ International School for Advanced Studies
(SISSA), Trieste, Italy.}

\date{\today} 

\maketitle 


\begin{abstract}

We investigate by numerical simulations and analytical calculations
the Bak-Sneppen model for biological evolution in scale-free networks.
By using large scale numerical simulations, we study the avalanche
size distribution and the activity time behavior at nodes with
different connectivities.  We argue the absence of a critical barrier
and its associated critical behavior for infinite size systems. These
findings are supported by a single site mean-field analytic treatment
of the model.

\end{abstract}

Many real systems, ranging from biological systems such as food webs
\cite{will00,bha99,jeo00} to communication systems\cite{alb99,fal99}, 
exhibit properties that lies in between those of regular lattices 
and random graphs\cite{bol85}. 
They are usually referred to as complex networks \cite{str01}. 
These networks may have large clustering coefficients like regular 
lattices but also have a small diameter which is a typical 
feature of random graphs. 
Among the class of complex networks, a particular role is played by 
{\em scale-free} networks (SF)\cite{bar99} in which there 
are not characteristic fluctuations in the connectivity of the nodes.
This implies that the  probability $P_k$ that a node is connected
with other $k$ nodes follows a power law, {\em i.e.}, $P_k\sim
k^{-\gamma}$, in contrast to  {\em exponential} graphs in which
$P_k$ is exponentially bounded \cite{bol85,str01,ama00}. 

SF networks have been recently recognized to describe several real
growing networks \cite{jeo00,alb99,fal99,bar99,mon00}, and at the same
time have proved to show very peculiar features with respect to
physical properties such as damage tolerance\cite{alb00}, epidemic
spreading\cite{ves01}, and diffusion properties\cite{hub01}. It is
then natural to ask whether and to what extent the topology of these
complex networks would affect many of the results obtained for
punctuated evolution models in regular lattices.

In this Letter, we study the Bak-Sneppen model (BS) \cite{bak93} on SF
networks. We perform large scale numerical simulations and find that
contrary to what is observed in regular lattices and in exponential
networks, the system self-organizes into a stationary state
characterized by the lack of a critical threshold barrier in the
thermodynamic limit. This result is confirmed analytically by
constructing the geanological tree of the avalanches and performing a
mean-field approach which takes into account the strong fluctuations
of the network's connectivity distribution. Finally, we discuss the
consequences that the present results could have in the general
context of punctuated evolution models.

The standard BS model \cite{bak93} can be considered as an ecological
system formed by many species that interact one with each other if
they are in contact, {\em i.e.}, the interactions are local.  To each
node of the graph we allocate a random {\em fitness} barrier $b_i\in
[0,1)$ ($i=1,\dots,N$) that represents the ability of species to
survive or mutate \cite{bak93}. The fitness are initially uniformly
distributed between 0 and 1 and the dynamics is updated according to
the following rules: {\em i)} at each time step, the species with the
minimum barrier $b_{min}$ is located and mutated by assigning a new
random value for its fitness. This somehow mimics the Darwinian
principle that the least fit species evolve by mutation or disappear;
and {\em ii)} the species directly linked to the species with
$b_{min}$ change their fitnesses to new random numbers as the result
of their interactions. By applying these rules repeatedly, after a
transient period and regardless of the initial distribution of fitness
barriers, the fitness distribution $P(b)$ of the system evolve toward
a self-organized stationary state where several physical quantities
can be measured. In previous studies on {\em regular geometries} and
{\em exponential networks} \cite{fly93,boe96,chr98}, it has been found
that the distribution of barriers tends, in the limit of infinite
system sizes, into a step-like function characterized by the existence
of a single parameter $b_c$. In these cases, for $b<b_c$ the
distribution $P(b)=0$, otherwise, $P(b)$ is equal to a constant value.

\begin{figure}[b]
\begin{center}
\epsfig{file=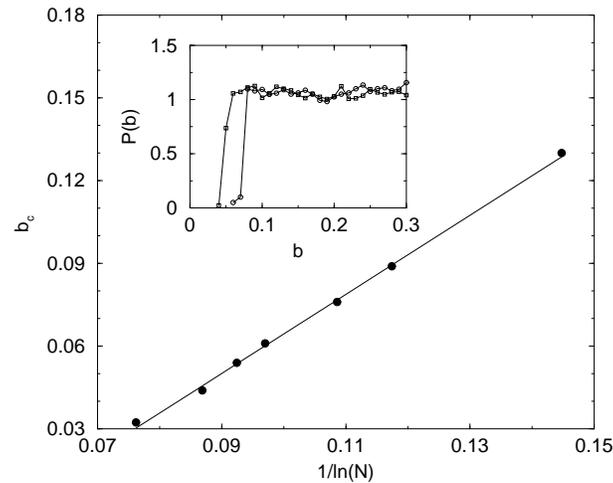,width=6.5cm,angle=-90,clip=1}
\end{center}
\caption{Threshold barriers for several system sizes ranging from
$N=10^3$ to $N=5\times 10^5$. The inset shows the
distribution of barriers in the stationary state for systems of size
$10^4$ (circles) and $10^5$ (squares) for $b<0.3$.}
\label{figure1}
\end{figure}

\begin{figure}[b]
\begin{center}
\epsfig{file=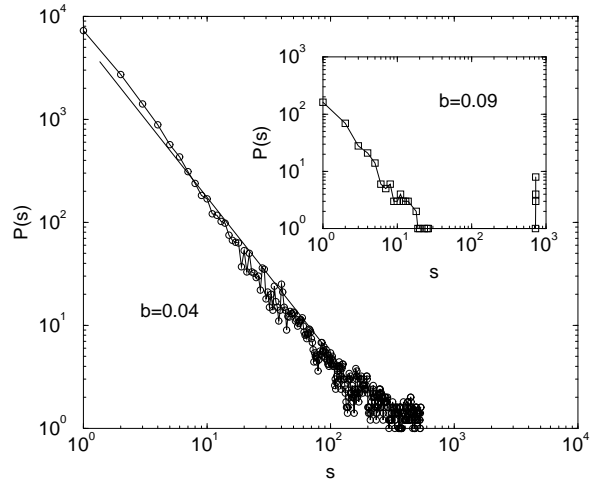,width=6.5cm,angle=-90,clip=1}
\end{center}
\caption{Avalanche size distributions for $b=0.04$ and
$b=0.09$ (inset). The system, which consists of $N=10^5$ species, has
the threshold value at $b_c=0.044$. The data were recorded after a
transient of $10^6$ mutations. A full line corresponding to a power
law with exponent $\tau=1.5$ has been drawn for comparison.}
\label{figure2}
\end{figure}

In order to study the BS model on a SF topology we use the network
obtained using the algorithm of Ref.\cite{bar99}. This is a stochastic
growing network model in which at each time step a new node (or
vertex) is added to the network and connected preferentially to the
already existing ones with a probability that depends on their
connectivities.  In practice, we start with a small number $m_0$ of
disconnected nodes and at each time step the network grows by adding a
new vertex. This is connected to $m$ old ones $i$ with a probability
$\Pi(k_i)=k_i/\sum_j k_j$. By iterating this scheme, a network of size
$N$ with connectivity distribution $P_k\sim k^{-3}$ develops.  As a
first step, we performed numerical simulations of the BS model in SF
networks with sizes ranging from $N=10^3$ to $N=5\times 10^5$. The BS
dynamics is iterated to achieve a stationary state giving rise to the
step function behavior usually observed in regular lattice. On SF
networks, however, it appears that the critical barrier $b_c$ is not
an intrinsic quantity and that $b_c\to 0 $ if $N\to \infty$.  Figure\
\ref{figure1} shows the dependency of the critical barrier with the
system size. It turns out that as the system size is increased the
critical barrier $b_c$ shifts leftward and goes to zero
logarithmically as $b_c\sim 1/ln N$. Thus, in the thermodynamic limit
($N\rightarrow\infty$), there is no threshold barrier;
i.e. $b_c=0$. Above the threshold, the distribution of barriers is
flat with overimposed statistical fluctuations.  A further check of
this behavior is provided by the analysis of the burst activity
(avalanche) in the system.  An avalanche is defined as the number of
subsequent mutations below a certain threshold. In Fig.\
\ref{figure2}, we show the avalanche size distribution for two
different values of avalanche threshold $b$. For this system, which
consists of $N=10^5$ species, the critical barrier is $b_c=0.044$. For
values of $b$ below $b_c$, the avalanches are distributed according to
a power law $P(s)\sim s^{-\tau}$ with $\tau=1.55\pm 0.05$, a value
close to the mean field exponent $\tau=3/2$ observed in other versions
of the BS model \cite{boe96} and in self-organized critical models
\cite{vesp98}. The inset illustrates that as soon as the $b$-barrier
is placed above the threshold, the distribution splits into two parts
(see the inset) and is characterized by the excess of large avalanches
signaling that we are in the supercritical region. If the barrier $b$
were further increased, then there would be only one avalanche whose
size will be limited only by the observation time. Besides, for very
large system sizes the absence of a critical barrier makes that even
for small $b$ the avalanche size distribution looks like the inset of
Fig\ \ref{figure2} indicating the lack of a critical point.

The numerical evidence tells us that the BS model behavior radically
changes in SF networks. The lack of a threshold can be intuitively
understood by recalling that for regular lattices and for networks in
which the connectivity distribution is exponentially bounded, the
threshold barrier decreases as the number of neighbors with which a
species interacts increases. This is the case studied in \cite{chr98},
where an ecology consisting of $N$ species, each one interacting with
$z_i$ neighbors, was studied.  In this model, the $z_i$ neighbors are
drawn from a Poisson distribution with a mean $\langle z
\rangle$. When increasing $\langle z \rangle$ and thus the
connectivity, it was observed that the threshold barrier $b_c$
decreases. In SF networks like the one studied here, the fact that the
fluctuations in the number of neighbors of each species diverge
($\langle k^2 \rangle=\infty$) makes null the threshold barrier when
$N\rightarrow \infty$.  It is worth remarking that the absence of a
critical threshold due to the strong fluctuations of the SF networks
connectivity has been recently reported for epidemic
spreading\cite{ves01,ym01} and percolation-like phase transitions
\cite{alb00}.

In order to support with analytical considerations  the
absence of a critical barrier, we will analyze the burst like activity
observed in the evolution process following a MF approach.
MF approaches are expected to give the right solution for 
growing random networks which are defined by  non-local random topologies.  
Let us define a  $b$ avalanche when there is one specie $i$ with $b_i\leq b$
while for all the others $b_j>b$. The $b$ avalanche 
lasts until all the barriers are above the $b$ barrier. 
Thus, if there is a critical barrier $b_c$, the avalanches should 
always be finite for $b<b_c$ since below
the threshold it is expected that no species remain forever. Otherwise,
there is a nonzero probability to observe an infinite avalanche that
approaches one as we move away from the critical barrier. Let us
assume that $0<b<b_c$. Thus, a $b$ avalanche must be finite.

Within a $b$ avalanche one can build a geanological tree as it was
shown in Refs \cite{boe96}. A node in this tree represents a species
whose barrier goes below $b$ at some step. Since we have assumed that
the avalanche is finite this species will be selected for evolution at
some later step, assigning a new random barrier to it and its
neighbors. In this process, the barrier of the species itself and
those of its neighbors may go below the threshold again and, in such a
case, they are represented by other nodes in the tree. The fact that
these nodes are causally related is represented by a direct link from
the ancestor (the species with the minimum barrier) to its sons (the
species whose barriers go below $b$). The generation $t$ of a node in
this tree is then defined as the number of nodes one needs to pass in
order to arrive to that node starting from the node that triggered the
avalanche. As we have assumed that the avalanche is finite, the relative
density of species whose fitness values are below the critical barrier
should vanish in the stationary state.

Let $\phi_k(t)$ be the relative density of nodes at generation $t$
with given connectivity $k$. This connectivity is taken as the
connectivity of the real SF network, where the BS evolution rules take
place, and not to that corresponding to the geanological tree. The
rate equations for $\phi_k(t)$ are given by
\begin{equation}
\partial_t\phi_k(t)=-(1-b)\phi_k(t)+bk[1-\phi_k(t)]\Theta(b).
\label{eq:t1}
\end{equation}
The first term on the r.h.s. takes into account that, when a species
is selected for evolution, with probability $1-b$ its new randomly
assigned barrier may be above $b$. The second term considers the
fraction of nodes $1-\phi_k(t)$ whose barriers are above $b$, but
change their barriers because they have a link to the species selected
for evolution. The factor $b$ takes into account that the new barrier
will be below $b$ with probability $b$. $\Theta(b)$ denotes the
probability that a link points to a node selected by evolution, and
$k\Theta(b)$ the probability that a species with connectivity $k$ have
a link pointing to the species which are going to evolve. 
By using the MF  considerations  of Ref.\cite{ves01},  this probability is
given by the number of links belonging to species with connectivity
$k$, $kP_kN$, divided by the total number of links
$\sum_ssP_sN$. Hence, making the summation over $k$ it results that
\begin{equation}
\Theta(b)=\sum_k\frac{kP_k\phi_k}{\sum_ssP_s}.
\label{eq:t3}
\end{equation}
In the stationary state $\partial_t\phi_k(t)=0$ and from Eq.\
(\ref{eq:t1}) and (\ref{eq:t3}) we obtain two self-consistent
equations from which one obtains $\Theta(b)$ and $\phi_k$. Finally,
the stationary density of nodes with barrier below $b$ is
$\phi=\sum_k\phi_k$. This set of equations is analogous to the one
studied in \cite{ves01}. The solution gives that for SF networks with
$2<\gamma\leq3$ and for any value of $b$ the stationary density of
species below $b$ is finite and, therefore, in the thermodynamic limit
there is a finite probability to obtain an infinite avalanche where
strictly speaking, $\phi$ is nonzero. However, this result is in
contradiction with our initial assumption that $0<b<b_c$, for which
the avalanches should be finite. Hence, we conclude that for
$2<\gamma\leq3$ there is not finite threshold $b_c$ when
$N\rightarrow\infty$. For any finite system size, the threshold,
although very small, is not zero since there is a finite probability
that all species have their fitness values above the critical barrier
at the same time. It is worth noting at this point that the above
arguments can be used to easily show that for exponential networks
with an average connectivity $\langle k \rangle$ the scenario at the
stationary state is completely different. In this case
$\Theta(b)\approx\phi$ and it results that there is a finite threshold
barrier which is given by $b_c=\frac{1}{1+\langle k \rangle}$, as can
be seen by direct comparison with the numerical values reported in
\cite{chr98}, $b_c=0.3446\approx 1/3$ and $b_c=0.2575\approx 1/4$ for
$\langle k \rangle=2$ and $\langle k \rangle=3$, respectively. Thus,
for exponential networks, the threshold barrier is in general finite
and it is determined by the first moment of the degree distribution
$\left< k\right>$. On the contrary, in SF networks, the lack of a
threshold barrier takes place when the second moment $\left<
k^2\right>$ diverges, {\em i.e.} when $\gamma\leq3$. The peculiar
topology of these networks gives rise to the existence of a few nodes
with a very high number of neighbors, and thus the fluctuations in the
number of neighbors emanating from each node are unbounded (that is,
the second moment diverges). This fact becomes determinant for the
dynamics of the system. We would like to remark that our results are
on the same line as those obtained in spreading dynamics on SF
networks where the critical parameter is related to the first and the
second moments through the relation $\langle k \rangle/ \langle k^2
\rangle$ \cite{ves01,ym01} which implies a vanishing threshold in the
thermodynamic limit. However, we emphasize that real networks have
always a finite size $N$ and thus a finite effective threshold,
although extremely small for very large system sizes.

\begin{figure}[b]
\begin{center}
\epsfig{file=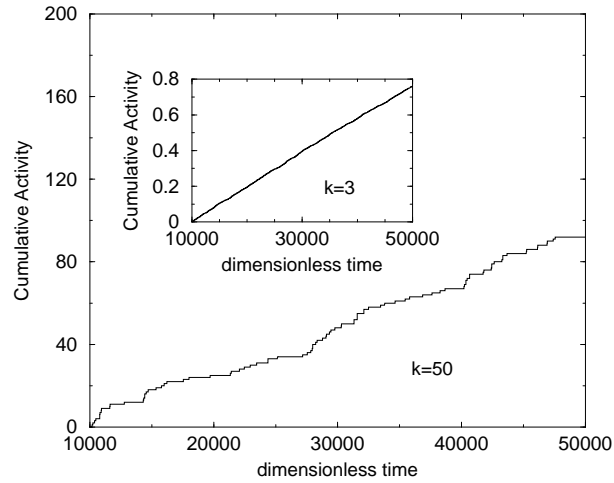,width=6.5cm,angle=-90,clip=1}
\end{center}
\caption{Cumulative activity for species with $k=50$ and $k=3$ (inset)
neighbors. In this case, $k=3$ corresponds to the minimally connected
species. The activity strongly depends on the species'
connectivity and for highly connected species it is $\sim$ 100 times
that of the minimally connected sites. The punctuated equilibrium
behavior can be nevertheless observed for both sub-populations.}
\label{figure3}
\end{figure}

The peculiar features of the BS model depend on the highly
heterogeneous nature of SF networks. As a further evidence of this
heterogeneity, we explore how the activity in the network is
distributed according to the nodes' connectivity.  We have measured
the number of times $n_k$ a node with connectivity $k$ has been
selected to mutate because it has the minimum barrier among all the
species. The results obtained shows that the activity is concentrated
in the species with high connectivities. This indicates that the
activity patterns are strongly correlated with the connectivity of
each site. Species with high degrees of interaction with their
neighborhoods are thus more susceptible to mutate or become extinct
than their counterparts.  Speciation mechanisms \cite{may70} during
evolution have been suggested as key natural processes to explain the
evolutionary patterns found in all lineages. For speciation to be
accomplished, it has been proposed both geographical isolation and
environmental stress as important triggering factors. In the model
studied here, the nodes in the network represent species, whereas the
links between them correspond to the complex dependency relationships
that can be established in an ecosystem, for example, through a food
web or predator-prey relationship. Therefore, the species with high
connectivities may be thought of as being more stressed than others
with low connectivities are. Consequently, we expect that the
punctuated equilibrium pattern be manifested in a significant
reduction in the cumulative activity of the lowly-connected species as
compared to that of the highly-connected ones. Figure\ \ref{figure3}
shows the cumulative activity for species with connectivity $k=50$ and
$k=3$ (inset). As can be clearly seen, the punctuated equilibrium
behavior appears in both sub-populations. However, the higher the
species connectivities are, the shorter the periods of intensive
activity are. Thus, for the {\em same} time window, speciation and
rapid episodes of major evolutionary changes are expected to take
place in the more stressed, highly dependent species. This correlation
has been observed in several studies \cite{eld89}. Besides, the fact
that speciation can occur rapidly over a large area containing
millions of individuals has been documented by Williamson
\cite{will81}, who studied a series of mollusk fauna of the eastern
Turkana basin, in Africa.

In summary, we have studied the BS model on scale-free networks. We
have shown that the highly interacting species play a key role in the
evolution process. Numerical results as well as analytical arguments
point out the absence of a critical threshold barrier in the
thermodynamic limit. For real system, which are always of finite size,
the threshold, although very small, exists. We have also found that
the activity patterns are strongly correlated with the topology of the
network. Finally, we have outlined the possible implications our
results may have for speciation events.

We are grateful to A. Vespignani for helpful comments and discussions.

\end{document}